\documentclass{IEEEtran}
\usepackage{cite}
\usepackage{amsmath,amssymb,amsfonts}
\usepackage{multirow}
\usepackage{algorithmic}
\usepackage{amsfonts,cite,times,url,hyperref,bm}
\usepackage{graphicx}
\usepackage{textcomp}
\usepackage{color}
\usepackage{circuitikz}

\newtheorem{assumption}{Assumption}
\usepackage{balance}

\def\BibTeX{{\rm B\kern-.05em{\sc i\kern-.025em b}\kern-.08em
    T\kern-.1667em\lower.7ex\hbox{E}\kern-.125emX}}
\begin{document}
\title{Reachability-based Time-domain Distance Protection}
\author{Joshua A. Taylor, \IEEEmembership{Senior Member, IEEE}, Nathan Baeckeland, \IEEEmembership{Member, IEEE}, and Alejandro D. Dom\'inguez-Garc\'ia, \IEEEmembership{Fellow, IEEE}
\thanks{This work was supported in part by the National Science Foundation under Grant 2411925. This work was authored in part by the National Laboratory of the Rockies for the U.S. Department of Energy (DOE), operated under Contract No. DE-AC36-08GO28308. The views expressed in the article do not necessarily represent the views of the DOE or the U.S. Government. The U.S. Government retains and the publisher, by accepting the article for publication, acknowledges that the U.S. Government retains a nonexclusive, paid-up, irrevocable, worldwide license to publish or reproduce the published form of this work, or allow others to do so, for U.S. Government purposes.}
\thanks{Josh A. Taylor is with the Department of Electrical and Computer Engineering,
       New Jersey Institute of Technology, Newark, NJ, USA. E-mail:
        {\tt\small jat94@njit.edu}.}
\thanks{Nathan Baeckeland is with the National Laboratory of the Rockies,
       Golden, CO, USA. E-mail:
        {\tt\small Nathan.Baeckeland@nlr.gov}.}
\thanks{Alejandro D. Dom\'inguez-Garc\'ia is with the Department of Electrical and Computer Engineering, University of Illinois at 
	Urbana-Champaign, Urbana, IL, USA. E-mail:
        {\tt\small aledan@illinois.edu}.}
}

\maketitle

\begin{abstract}
Distance relays detect faults on transmission lines from local voltage and current measurements. In this paper, we characterize the set of time-domain measurements a distance relay can observe during a fault as a reachable set. We define corresponding fault tests in terms of set-based state estimation. This allows us to model the full network as an RLC circuit with voltage and current sources. In general, computing reachable sets and set-based state estimates is intractable. For each fault type, we construct a reduced-order, two-dimensional model in terms of the apparent voltage and current seen by the relay. This makes all computations efficient, aligns with standard quantities in distance protection, and allows us to define instantaneous fault tests in terms of unsolved differential equations.
\end{abstract}


\section{Introduction}

Distance relays detect faults on transmission lines. Traditionally, this has meant computing voltage and current phasors, dividing to get an apparent impedance, and then checking if the impedance is inside of a certain set in the complex plane, known as a characteristic~\cite{anderson2021power}. Due to the need to compute phasors, detection typically occurs within 1.5 cycles. Time-domain protection schemes, rather than computing phasors, test if the relay's measurements are consistent with a differential equation~\cite{johns1995digital,schweitzer2015speed}. Time-domain schemes can detect faults in as fast as a quarter of a cycle and have shown promise for grids with inverter-based resources (IBRs) (cf.~\cite{adhikari2021source,kasztenny2022distance,baayeh2025fast,quintero2025enhanced,kasztenny2025locating}).

In this paper, we pose time-domain fault detection in terms of reachability~\cite{girard2005zono,althoff2021set} and set-based state estimation~\cite{schweppe1968recursive}. In Section~\ref{sec:model}, we model the network as a lumped parameter RLC circuit in which the sources are synchronous machines (SGs) and IBRs, and the fault, as usual, is a resistor. A relay on a given line faces uncertainty about a fault's location and resistance and the current flowing into the remote end of the line, which is uncertain because the relay does not observe the states or sources elsewhere in the network.

Given the set of possible initial states, the voltages and currents the relay can observe in the time following a fault comprise a reachable set. Each type of fault corresponds to a different system model and thus reachable set. Given a new measurement, we can refine each reachable set to obtain a set-based state estimate, from which new reachable sets can be computed, and so on. In Section~\ref{sec:test}, we define a fault test in which the relay checks if each new measurement is in the corresponding reachable set, or, equivalently, if the new set-based state estimate is nonempty. As stated, this test is impractical for two reasons: the relay has limited information about the sources in the network, and computing reachable sets and set-based state estimates is hard.

The need for information elsewhere in the network can be somewhat alleviated by using incremental quantities, in which the voltages and currents from one or more cycles earlier are subtracted off of new measurements~\cite{benmouyal2001combined,schweitzer2015speed}. If the sources remain periodic shortly after faults, then the relay need only know the pre-fault voltage at the fault point, which can be computed from its earlier measurements. In Section~\ref{sec:incremental}, we derive the dynamics of the incremental model of the full network, which does depend on the structure of the network, but not the present values of the sources.

There is an extensive literature on tractable approximations of reachable sets and set-based state estimates via, e.g., ellipsoids~\cite{schweppe1968recursive}, parallelotopes~\cite{chisci1996recursive}, zonotopes~\cite{alamo2005guaranteed}, and constrained zonotopes~\cite{scott2016constrained}. These are typically outer approximations that always contain the corresponding set, but can also be conservative. Here, in Section~\ref{sec:rom}, we instead construct a reduced-order model for each fault type in which the two states are the corresponding apparent current and voltage. While this does not result in an outer approximation, it simplifies the problem in several ways: the system matrices can be computed offline; the states are standard quantities in distance protection; and the set-based state estimate is the relay's observation, eliminating the need to take intersections.

In Section~\ref{sec:instant}, we define instantaneous fault tests, in which the relay checks if the apparent voltage and current it measures are consistent with the unsolved, reduced-order system model. This is more efficient because the differential equation need not be solved, and it has similar form to existing time-domain schemes~\cite{schweitzer2015speed,adhikari2021source,kasztenny2025locating}. In Section~\ref{sec:ex}, as proof of concept, we test the instantaneous test on the
Simulink model from~\cite{baeckeland2025unified}, which is based on the IEEE 14-bus test system.

Our approach is novel in that it is the first to make use of reachability, set-based state estimation, and model order reduction in this context. The novelty is significant because the set of measurements a relay can take during a fault are, by definition, a reachable set; and the natural fault test is an instance of set-based state estimation. We also differ from existing schemes in that we keep entirely in the time-domain, whereas most make heuristic use of phasor impedances. A notable exception is \cite{adhikari2021source}, which, as here, models the line as $\pi$-equivalent circuit, but does not model the network and uses communication to handle the remote infeed. This paper is also related to our earlier work on distance protection in the phasor domain~\cite{taylor2025geometry,taylor2025phasor}, in which we used the Minkowski sum and zonotopes to account for the network and its uncertainty.




\section{Modeling}\label{sec:model}

Let $\mathcal{B}$ be the set of buses and $\mathcal{L}$ the set of lines in the network. The voltage at bus $k\in\mathcal{B}$ is denoted $v_k\in\mathbb{C}^3$. The current from bus $k$ to $l$, $kl\in\mathcal{L}$, is denoted $i_{kl}\in\mathbb{C}^3$. We use superscript a, b, or c to indicate the phase.

Suppose there is a physical line connecting bus L to R and a relay at bus L. We refer to L as the local bus and R as the remote bus. A fault occurs at time $t=0$ at location on the line that we model as a virtual bus, F. Note that in this setup, both before and after the fault, $\textrm{LF}\in\mathcal{L}$ and $\textrm{RF}\in\mathcal{L}$, and $\textrm{LR}\notin\mathcal{L}$. There are eleven types of faults: line-to-ground (LG), line-to-line (LL), double line-to-ground (LLG), and the symmetrical faults, abc and abcg. Each scenario has a model, which we index with the set
\[
\mathbb{F} = \{
\underset{\textrm{LG}}{\underbrace{\textrm{ag},\textrm{bg},\textrm{cg}}},
\underset{\textrm{LL}}{\underbrace{\textrm{ab},\textrm{ac},\textrm{bc}}},
\underset{\textrm{LLG}}{\underbrace{\textrm{abg},\textrm{acg},\textrm{bcg}}},
\underset{\textrm{symmetrical}}{\underbrace{\textrm{abc},\textrm{abcg}}}
\}.
\]
In all cases, the fault's maximum resistance is $R_{\textrm{F}}$. The fault's normalized location and resistance are $m_{\textrm{T}}\in\left[0,1\right]$ and $m_{\textrm{F}}\in[0,1]$. We let $m=(m_{\textrm{T}},m_{\textrm{F}})$.

The relay measures $v_{\rho}(t)$ and $i_{\rho}(t)$, which we define in terms of the network model in Appendix~\ref{sec:app:relobs}. For each $\eta\in\mathbb{F}$, the relay computes an apparent voltage, $v_{\textrm{A}}^{\eta}(t)$, and current, $i_{\textrm{A}}^{\eta}(t)$. For example, for an ab fault, $v_{\textrm{A}}^{\textrm{ab}}(t) = v_{\rho}^{\textrm{a}}(t)-v_{\rho}^{\textrm{b}}(t)$ and $i_{\textrm{A}}^{\textrm{ab}}(t) = i_{\rho}^{\textrm{a}}(t)-i_{\rho}^{\textrm{b}}(t)$. For each $\eta\in\mathbb{F}$, let $\psi^{\eta}$ be such that $i_{\textrm{A}}^{\eta}(t)=\psi^{\eta}i_{\rho}(t)$ and $v_{\textrm{A}}^{\eta}(t)=\psi^{\eta}v_{\rho}(t)$. For example $\psi^{\textrm{ag}}=\begin{bmatrix}1&0&0\end{bmatrix}$ and $\psi^{\textrm{ab}}=\begin{bmatrix}1&-1&0\end{bmatrix}$.

We want to know the set of all possible $i_{\textrm{A}}^{\eta}(t)$ and $v_{\textrm{A}}^{\eta}(t)$ the relay can measure when there is a fault of type $\eta\in\mathbb{F}$ on its line. We define this in terms of reachability and set-based state estimation in Section~\ref{sec:test}.


\subsection{Dynamics}\label{sec:dynamics}

The state vector, $x(t)$, consists of line currents, nodal voltages, and potentially the internal states of the IBRs and SGs. The input vector, $u(t)$, contains voltage and current sources representing the SGs, IBRs, and potentially loads. We write out the model explicitly in Appendix~\ref{app:dynamics}.

We model the network as an LTI system that changes after a fault. The dynamics before the fault ($t<0$) are
\begin{subequations}
\label{eq:ss}
\begin{align}
\frac{dx(t)}{dt} &= A_mx(t) + Bu(t).\label{eq:ss:pre}
\end{align}
After a fault ($t\geq0$) of type $\eta\in\mathbb{F}$, the dynamics are
\begin{align}
\frac{dx(t)}{dt} &= A_m^{\eta}x(t) + Bu(t).\label{eq:ss:post}
\end{align}
The equations that specify $A_m$, $A_m^{\eta}$, and $B$ are given in Appendix~\ref{app:dyn}. Note that the state vector and inputs are the same in (\ref{eq:ss:pre}) and (\ref{eq:ss:post}). Also note that the state transition matrices, $A_m$ and $A_m^{\eta}$, depend on $m$; before a fault, the dynamics depend on the location of the virtual bus, $m_{\textrm{T}}$, and after a fault, they depend on both the fault's location and resistance, $m_{\textrm{F}}$.

The output for fault type $\eta\in\mathbb{F}$ is the corresponding apparent voltage and current,
\[
y^{\eta}(t) = \begin{bmatrix}
v_{\textrm{A}}^{\eta}(t)\\i_{\textrm{A}}^{\eta}(t)
\end{bmatrix}\in\mathbb{R}^2.
\]
These are standard quantities in classical fault loop analysis and leads to simple, two-dimensional computations. We write this generically as
\begin{align}
y^{\eta}(t) &= H^{\eta}x(t).
\end{align}
\end{subequations}
We note that there are other a potentially more informative choices of observation, e.g., $\begin{bmatrix}
v_{\rho}(t)&i_{\rho}(t)
\end{bmatrix}^{\top}\in\mathbb{R}^6$.

\subsection{Incremental model}\label{sec:incremental}

For a given quantity $q(t)$, the corresponding incremental quantity is
\[
\tilde{q}(t)=q(t)-q(t-p\delta),
\]
where $\delta$ is the duration of one cycle and $p$ a positive integer. We hereon set $p=1$.
\begin{assumption}\label{a:inc}
For all $t\in[0,\delta]$, $u(t)=u(t-\delta)+\lambda$.
\end{assumption}
$\lambda$ is constant source drift, which we detail in Section~\ref{sec:uncertainty}. If we assume that all sources are periodic in the time shortly before and after a fault, then incremental quantities do not depend on the operating point of the sources (cf.~\cite{benmouyal2001combined,schweitzer2015speed}).

For each $\eta\in\mathbb{F}$, let
\[
\hat{A}^{\eta}_m  = A^{\eta}_m - A_m.
\]
The dynamics of the incremental system over $t\in[0,\delta]$ are
\begin{subequations}
\label{eq:inc}
\begin{align}
\frac{d\tilde{x}(t)}{dt} &= A_m^{\eta}x(t) - A_mx(t-\delta) + B(u(t)-u(t-\delta))\nonumber\\
&= A_m^{\eta}x(t) - A_m^{\eta}x(t-\delta) + \hat{A}_m^{\eta}x(t-\delta) - B\lambda \nonumber\\
&= A_m^{\eta}\tilde{x}(t) + \hat{A}_m^{\eta}x(t-\delta) - B\lambda.\label{eq:incdyn}
\end{align}
The inputs are the prior value of the state, $x(t-\delta)$, and the drift, $\lambda$. The output of the incremental system is
\begin{align}
\tilde{y}^{\eta}(t) &= H^{\eta}\tilde{x}(t).\label{eq:out}
\end{align}
\end{subequations}


$\hat{A}^{\eta}_m x(t-\delta)$ is a vector of all zeros except for the rows corresponding to $\tilde{v}_{F}(t)$.  These rows of $\hat{A}^{\eta}_m$ are nonzero only in columns corresponding to $v_{\textrm{F}}(t-\delta)$. From this we can see that the incremental system depends on the network structure and, through $B$, how the sources are modeled, but not $u(t)$. Unlike (\ref{eq:ss}), its only input is the pre-fault voltage at the fault point, $v_{\textrm{F}}(t-\delta)$, and potentially $\lambda$. We show how to calculate $v_{\textrm{F}}(t-\delta)$ from $v_{\textrm{L}}(t-\delta)$ and $i_{\textrm{L}}(t-\delta)$ in Appendix~\ref{sec:app:relobs}.


\subsection{Uncertainty}\label{sec:uncertainty}

There are several sources of uncertainty in (\ref{eq:inc}). A fault's location and resistance, $m=(m_{\textrm{T}},m_{\textrm{F}})$ are unknown. Let $\mathcal{M}=\left[0,1\right]\times [0,1]$. Then $m\in\mathcal{M}$.

The relay has limited information about the network. The initial state, $\tilde{x}_0$, has uncertainty set $\mathcal{X}_0$, and the source drift, $\lambda$, has uncertainty set $\Lambda$. If the system was in steady state before the fault, then $\mathcal{X}_0=\{\bm{0}\}$, and if the sources are perfectly periodic, $\Lambda=\{\bm{0}\}$, which simplifies our setup. Otherwise, $\mathcal{X}_0$ and $\Lambda$ add conservatism. In this case, for tractability, we model both as zonotopes~\cite{ziegler1995lectures}.

Note that all uncertain quantities in (\ref{eq:inc}) are constant for $t\in[0,\delta]$. Due to the uncertainty, the relay cannot know the exact state of the system. The state estimate, $\mathcal{X}^{\eta}_{\textrm{E}}(t)$, is the set of all possible values of $\tilde{x}(t)$ if there is a fault of type $\eta\in\mathbb{F}$. Note that $\mathcal{X}^{\eta}_{\textrm{E}}(0)=\mathcal{X}_0$. We describe the evolution of $\mathcal{X}^{\eta}_{\textrm{E}}(t)$ in the next section.

We use the following notation for uncertainty sets. Given a set $\mathcal{X}$ and appropriately dimensioned matrix $\Gamma$, we let
\[
\Gamma \mathcal{X} = \left\{
\Gamma x \;\left | \; x \in \mathcal{X} \right.
\right\}.
\]
The Minkowski sum of two sets, $\mathcal{X}$ and $\mathcal{Y}$, is
\[
\mathcal{X}\oplus\mathcal{Y}=\left\{x+y\;|\; x\in\mathcal{X},\;y\in\mathcal{Y}\right\}.
\]

\section{Fault tests}\label{sec:test}

For each fault type $\eta\in\mathbb{F}$, the relay observes the output (\ref{eq:out}) at discrete time instants $t_k$, $k\in\mathbb{N}$. We denote the observations $\breve{y}^{\eta}(t_k)$. The consistent state set is
\[
\mathcal{X}_{\textrm{O}}^{\eta}(t_k) = \left\{
\tilde{x}(t_k)\;|\; H^{\eta}\tilde{x}(t_k)=\breve{y}^{\eta}(t_k)
\right\}.
\]
For each $\eta\in\mathbb{F}$, the reachable set from time $t_{k-1}$ to $t_k$ is
\begin{align}
\mathcal{X}_{\textrm{R}}^{\eta}(t_{k-1},t_k) &= \left\{
\tilde{x}(t_k)\;|\;\tilde{x}(t)\textrm{ satisfies (\ref{eq:incdyn}) for } t\in[t_{k-1},t_k],\right.\nonumber\\
&\hspace{0mm} \left.\;m\in\mathcal{M},\;\lambda\in\Lambda,\;\tilde{x}(t_{k-1})\in\mathcal{X}_{\textrm{E}}^{\eta}(t_{k-1})
\right\}.\label{eq:reach}
\end{align}
An iteration of set-based state estimation is as follows.
\begin{enumerate}
\item Given $\mathcal{X}_{\textrm{E}}^{\eta}(t_{k-1})$, we obtain $\mathcal{X}_{\textrm{R}}^{\eta}(t_{k-1},t_k)$ from (\ref{eq:reach}).
\item Given $\breve{y}^{\eta}(t_k)$, the next state estimate is
\begin{align}
\mathcal{X}_{\textrm{E}}^{\eta}(t_k) &= \mathcal{X}_{\textrm{R}}^{\eta}(t_{k-1},t_k) \cap \mathcal{X}_{\textrm{O}}^{\eta}(t_k).\label{eq:seinter}
\end{align}
\end{enumerate}

The relay may test for faults as follows. Suppose that a fault of type $\eta\in\mathbb{F}$ might have occurred just after time $t_0$. If yes, then $\mathcal{X}_{\textrm{E}}^{\eta}(t_1)\neq\emptyset$. This is equivalent to $
\breve{y}^{\eta}(t_1)\in H^{\eta}\mathcal{X}_{\textrm{R}}^{\eta}(t_{0},t_1)$, i.e., the relay's next observation is consistent with the model of fault $\eta\in\mathbb{F}$. In each subsequent time period, $t_k$, the relay can check if
\begin{align}
\breve{y}^{\eta}(t_k)&\in H^{\eta}\mathcal{X}_{\textrm{R}}^{\eta}(t_{k-1},t_k).\label{eq:yinHR}
\end{align}
If (\ref{eq:yinHR}) is satisfied for $N\geq1$ time periods, the relay declares a fault of type $\eta\in\mathbb{F}$. The choice of $N$ is a tunable parameter that trades off speed and accuracy.

The sets $\mathcal{X}_{\textrm{R}}^{\eta}(t_{k-1},t_k)$ and $\mathcal{X}_{\textrm{E}}^{\eta}(t_k)$ are hard to compute. If the matrix $A^{\eta}_m$ did not depend on $m$, we could use any of several approximations~\cite{schweppe1968recursive,chisci1996recursive,alamo2005guaranteed,scott2016constrained}. While some are accurate and all have polynomial-time complexity, it is not clear that they are fast or simple enough for time-domain distance protection.


\section{Reduced-order approximation}\label{sec:rom}

To improve tractability, we construct approximate tests based on reduced-order systems with state $\tilde{y}^{\eta}(t)\in\mathbb{R}^2$, $\eta\in\mathbb{F}$. We write each reduced-order system as
\begin{align}
\frac{d\tilde{y}^{\eta}(t)}{dt}
&= \Omega_m^{\eta} \tilde{y}^{\eta}(t) + \hat{\Omega}_m^{\eta} v_{\textrm{F}}(t-\delta) - \chi_m^{\eta}\lambda.\label{eq:incdynred}
\end{align}
Obtaining the matrices $\Omega_m^{\eta}$, $\hat{\Omega}_m^{\eta}$, and $\chi_m^{\eta}$ is a structured model reduction problem. In Appendix~\ref{app:singular}, we use singular perturbation~\cite{kokotovic1999singular}, and note that other, potentially more accurate options exist, e.g., structured balanced truncation~\cite{sandberg2010extension} and Krylov subspaces~\cite{freund2003model,althoff2019reachability}.

Let $\mathcal{Y}^{\eta}_{\textrm{O}}(t_k)\in\mathbb{R}^2$, $\mathcal{Y}^{\eta}_{\textrm{R}}(t_{k-1},t_k)\in\mathbb{R}^2$ and $\mathcal{Y}^{\eta}_{\textrm{E}}(t_k)\in\mathbb{R}^2$ respectively denote the consistent state set, reachable set, and state estimate for the reduced-order system. We may use these quantities as in Section~\ref{sec:rom} to construct approximate fault tests. Specifically, we declare that there is a fault of type $\eta\in\mathbb{F}$ at time $t_k$ if $\mathcal{Y}^{\eta}_{\textrm{E}}(t_k)\neq\emptyset$ or, equivalently, if $\breve{y}^{\eta}(t_k)\in\mathcal{Y}^{\eta}_{\textrm{R}}(t_{k-1},t_k)$.

The reduced-order model simplifies the computations 
in the fault test in two ways. First, $\mathcal{Y}^{\eta}_{\textrm{O}}(t_k)=\left\{\breve{y}^{\eta}(t_k)\right\}$, i.e., the consistent state set is the measurement. This implies that either $\mathcal{Y}^{\eta}_{\textrm{E}}(t_k)=\left\{\breve{y}^{\eta}(t_k)\right\}$ or $\mathcal{Y}^{\eta}_{\textrm{E}}(t_k)=\emptyset$.
Second, being two-dimensional, $\mathcal{Y}^{\eta}_{\textrm{R}}(t_{k-1},t_k)$ is easy to approximate.

\subsection{Corner approximation}\label{sec:corner}

The dependence of the system matrices on $m$ and subsequent model reduction makes it difficult to analytically characterize the reachable sets. We now describe a numerical approximation of $\mathcal{Y}^{\eta}_{\textrm{R}}(t_{k-1},t_k)$ that makes use of the fact that $\mathcal{M}$ and the reachable sets we seek are in two dimensions. Specifically, we evaluate the reachable set for several fixed values of $m\in\mathcal{M}$ and then take the convex hull. 

Let
\begin{align*}
\tilde{y}_{1,m}^{\eta}(t_k)
&=  e^{-(t_k-t_{k-1})\Omega_m^{\eta}}\breve{y}^{\eta}(t_{k-1})\nonumber\\
&\quad+\int_{t_{k-1}}^{t_k}e^{(t_k-\tau)\Omega_m^{\eta}}\hat{\Omega}_m^{\eta} v_{\textrm{F}}(t-\delta)d\tau\\
\tilde{y}_{2,m}^{\eta}(t_k)
&= \left(I-e^{-(t_k-t_{k-1}) \Omega_m^{\eta}}\right)\left(\Omega_m^{\eta}\right)^{-1} \chi_m^{\eta}\lambda.
\end{align*}
Observe that we have made the dependence on $m$ explicit through the subscripts. The solution of the reduced-order system, (\ref{eq:incdynred}), at time $t_k$ is
\begin{align*}
\tilde{y}^{\eta}_m(t_k)
&=  \tilde{y}_{1,m}^{\eta}(t_k)+\tilde{y}_{2,m}^{\eta}(t_k).
\end{align*}
Let
\[
\mathcal{Y}_{2,m}^{\eta}(t_k) = \left\{
\tilde{y}_{2,m}^{\eta}(t_k)\;|\; \lambda\in\Lambda
\right\}.
\]
$\mathcal{Y}_{2,m}^{\eta}(t_k)$ is a zonotope, which can be precomputed.

Let $\hat{m}_k\in\mathcal{M}$, $k=1,...,n$. Each $\hat{m}_k$ is a potential realization of $m$. Let $\textrm{ch}(\cdot)$ denote the convex hull operator. Then
\[
\mathcal{Y}^{\eta}_{\textrm{R}}(t_{k-1},t_k)\approx \textrm{ch}\left\{\bigcup_{k=1}^n \left\{\tilde{y}_{1,\hat{m}_k}^{\eta}(t_k)\right\}\oplus \mathcal{Y}_{2,\hat{m}_k}^{\eta}(t_k)
\right\}.
\]
The Minkowski sum is easy to compute because it is between a point and a zonotope; specifically, we add $\tilde{y}_{1,\hat{m}_k}^{\eta}(t_k)$ to the center of $\mathcal{Y}_{2,\hat{m}_k}^{\eta}(t_k)$. A simple choice for the $\hat{m}_k$ is the four corners of $\mathcal{M}$.


Note that we could instead construct reduced-order approximations and corresponding fault tests in the coordinates $\begin{bmatrix}\tilde{v}_{\rho}(t)&\tilde{i}_{\rho}(t)\end{bmatrix}^{\top}\in\mathbb{R}^6$. This would similarly simplify (\ref{eq:seinter}), but the six-dimensional reachable sets might remain challenging to compute.

\subsection{Instantaneous tests}\label{sec:instant}
Because all states in the reduced-order model are observed, we do not need to solve for any unobserved states. This means that instead of testing if an observation, $\breve{y}^{\eta}(t_k)$, is consistent with a reachable set, we can instead test if it consistent with the unsolved differential equation.

For $k=1,...,n$, let
\begin{align}
\epsilon^{\eta}_m(t) &= -\frac{d\tilde{y}^{\eta}(t)}{dt}
+ \Omega_m^{\eta} \tilde{y}^{\eta}(t) + \hat{\Omega}_m^{\eta} v_{\textrm{F}}(t-\delta) - \chi_m^{\eta}\lambda.\label{eq:ROMeps}
\end{align}
$\epsilon^{\eta}_m(t)$ is the error of the model $\eta\in\mathbb{F}$ at time $t\in[0,\delta]$. Because the relay measures $\tilde{y}^{\eta}(t)$, we can directly evaluate $\epsilon^{\eta}_m(t_k)$ upon receipt of $\breve{y}^{\eta}(t_k)$.
Let
\[
\Upsilon_m^{\eta} = \left\{-
\chi_m^{\eta}\lambda\;|\;\lambda\in\Lambda
\right\}.
\]

Let $\hat{m}_k\in\mathcal{M}$, $k=1,...,n$, as in Section~\ref{sec:corner}. Define
\begin{align}
\Xi^{\eta}(t)=\textrm{ch}\left\{\bigcup_{k=1}^n \left\{\epsilon^{\eta}_{\hat{m}_k}(t)\right\}\oplus \Upsilon_{\hat{m}_k}^{\eta}
\right\}.\label{eq:instch}
\end{align}
$\Xi^{\eta}(t)$ is an estimate of the set of all possible model errors. We test for faults by checking if
\begin{align}
\bm{0}&\in\Xi^{\eta}(t).\label{eq:insttest}
\end{align}

We now discuss the tractability of (\ref{eq:insttest}), which must be repeatedly evaluated online. All of the matrices in (\ref{eq:ROMeps}) can be computed offline for each of $\hat{m}_k$. Evaluating (\ref{eq:instch}) consists of $n$ Minkowski sums and a convex hull in $\mathbb{R}^2$, which can be computed efficiently with the gift wrapping algorithm~\cite{jarvis1973identification}. If $\Lambda$ is zonotopic, the Minkowski sums can be computed in linear time. A particularly simple implementation is $\Lambda=\{0\}$ and the $\hat{m}_k$ the four corners of the unit square. In this case, there are no Minkowski sums and the convex hull calculation involves four points.







\section{Example}\label{sec:ex}

\begin{figure}
    \centering
    \includegraphics[scale=1]{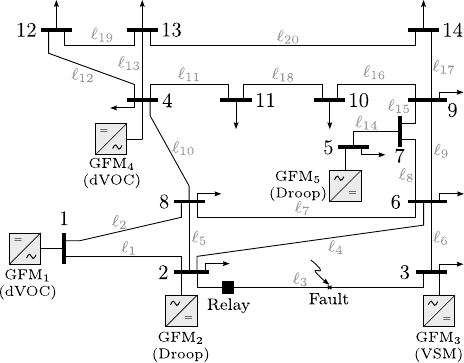}
    \caption{Modified IEEE 14-bus network~\cite{baeckeland2025unified} with $5$ GFM inverters. A phase a-to-phase b fault occurs on line $\ell_3$.}
    \label{fig:14bus}
\end{figure}

We simulate the instantaneous test of Section~\ref{sec:instant} on a Simulink model based on the IEEE 14-bus test system, as depicted in Figure~\ref{fig:14bus}. The system is fully described in~\cite{baeckeland2025unified}. The system features five grid-forming (GFM) inverters with a variety of primary-control methods; we use droop control, virtual synchronous machine, and dispatchable virtual oscillator control. Each of the GFM inverters has a current limiter to curtail the output during faults and grid disturbances. We subject the network to a phase-to-phase fault on line $\ell_3$. The relay is located at bus $2$. 


In the instantaneous test, we use a maximum resistance of $R_{\textrm{F}}=10\;\Omega$. We treat all IBRs as voltage sources and neglect their filter impedances. To ensure invertibility of the submatrix for model reduction, we added a small positive resistance to lines with zero resistance. We set $\Lambda=\{0\}$ and the $\hat{m}_k$'s to $[\nu,\nu],[1-\nu,\nu],[\nu,1-\nu]$, and $[1-\nu,1-\nu]$, where $\nu=10^{-4}$ to avoid division by zero.

\begin{figure}
  \includegraphics[scale=1.45]{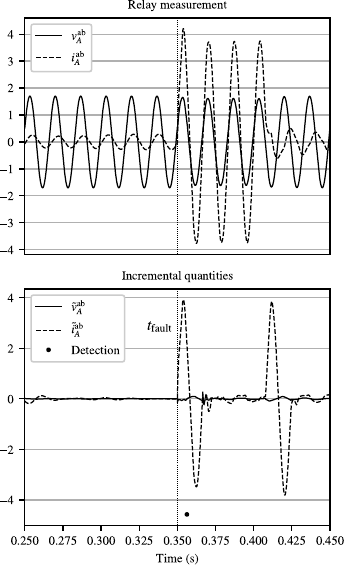}
  \caption{Low resistance ($0.01\;\Omega$) fault.}
  \label{fig:RLow}
\end{figure}

\begin{figure}
  \includegraphics[scale=1.45]{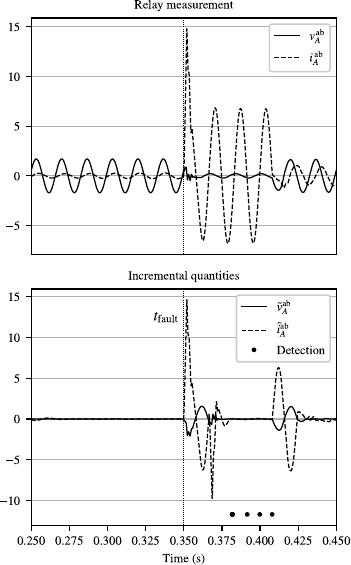}
  \caption{Medium resistance ($1\;\Omega$) fault.}
  \label{fig:RMid}
\end{figure}

\begin{figure}
  \includegraphics[scale=1.45]{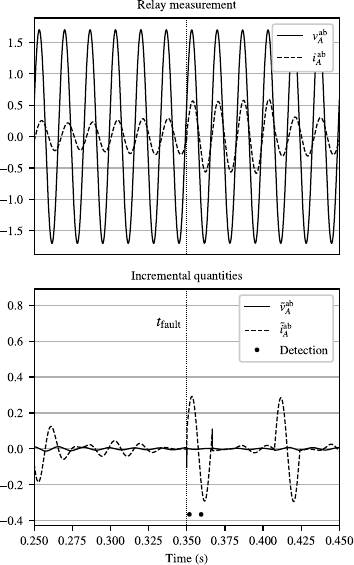}
  \caption{High resistance ($10\;\Omega$) fault.}
  \label{fig:RHigh}
\end{figure}

A phase a-to-phase b fault at the line's midpoint occurs at 0.35 seconds. We simulate the fault with resistances $0.01\;\Omega$ (Fig.~\ref{fig:RLow}), $1\;\Omega$ (Fig.~\ref{fig:RMid}), and $10\;\Omega$ (Fig.~\ref{fig:RHigh}). In each case, the figure shows the per unit voltage and current measured by the relay (top plot) and the per unit incremental voltage and current measured by the relay (bottom plot). In each bottom plot, a dot indicates detection by the instantaneous test.

In the high resistance case, detection occurs after 1/8 of a cycle. In this case, no current limiters are active and all IBRs behave like voltage sources. In the medium resistance case, some current limiters are active, leading to a slight drop in the voltage seen by the relay. Here detection occurs at roughly 3/8 of a cycle after the fault. In the low resistance case, all of the IBRs' current limiters are active, and detection occurs just after a full cycle.

\begin{figure}
    \centering
  \includegraphics[scale=1.45]{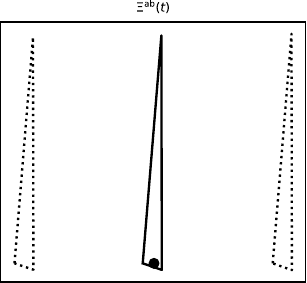}
  \caption{The set $\Xi^{\textrm{ab}}(t)$ before, during, and after detection. The dot denotes the origin.}
  \label{fig:RMidXi}
\end{figure}

For illustration, in Figure~\ref{fig:RMidXi}, we plot the set $\Xi^{\textrm{ab}}(t)$ for a $1\;\Omega$ fault just before, during, and just after detection occurred, with a dot at the origin.

\section{Conclusion}

In this paper, we have posed time-domain fault detection in terms of reachability and set-based state estimation. High-dimensional reachable sets and intersections are hard to compute, and it is not clear that even approximations will be practical on protection timescales. As a simple alternative, we have used model reduction to construct instantaneous tests, each of which entails checking the validity of a differential equation in the apparent voltage and current seen by the relay. Electromagnetic transient simulation has provided preliminary evidence that the tests are effective. Broadly, our work allows us to integrate network information into time-domain distance protection without sacrificing speed. 

We see several directions of future work. All of the tests in this paper are overreaching in that they aim to detect all possible faults on the line, and therefore potentially faults on neighboring lines. We intend to construct corresponding underreaching tests that might miss faults on the relay's line, but will not detect faults on neighboring lines. It is not clear what model is most appropriate for the IBRs, i.e., current sources, voltage sources, or a more complicated RLC circuit. On the other hand, it would be beneficial to reduce dependency on network modeling, e.g., so that schemes are more robust to setpoint and topology changes. Greater standardization of IBR fault responses may alleviate the need for network modeling (cf.~\cite{schweitzer2026consistency}), and other types of model reduction, e.g., Krylov subspace-based methods, might also prove less sensitive to the network.

\appendices

\section{System model}\label{app:dynamics}

We use a $\pi$-equivalent model of the lines. Note that this is an approximation of the telegrapher's equations, as we are not in steady state. Let $\mathcal{N}_k$ be the set of buses that are connected to $k$ by a line in $\mathcal{L}$. For $kl\in\mathcal{L}$, let $L_{kl}$, $C_{kl}$, and $R_{kl}$ be the line's inductance, charging capacitance, and resistance matrices, where the latter is diagonal. Let $L_{\textrm{LR}}$, $C_{\textrm{LR}}$, and $R_{\textrm{LR}}$ be the corresponding parameters of the physical line from LR. Recalling that $\textrm{LR}\notin\mathcal{L}$ and that the fault is a virtual bus, $\textrm{F}$, between buses $\textrm{L}$ and $\textrm{R}$. The parameters of line $\textrm{LF}$ are $L_\textrm{LF}=m_{\textrm{T}}L_{\textrm{LR}}$, $R_\textrm{LF}=m_{\textrm{T}}R_{\textrm{LR}}$, and $C_\textrm{LF}=m_{\textrm{T}}C_{\textrm{LR}}$, and similarly for $\textrm{RF}$, but multiplied by $1-m_{\textrm{T}}$. Let
\[
C_k(m)=\frac{1}{2}\left\{\begin{array}{ll}
m_{\textrm{T}}C_{\textrm{LR}}+\sum_{k\in\mathcal{N}_{\textrm{L}}\setminus\textrm{F}}C_{\textrm{L}k} & k=\textrm{L}\\
(1-m_{\textrm{T}})C_{\textrm{LR}}+\sum_{k\in\mathcal{N}_{\textrm{R}}\setminus\textrm{F}}C_{\textrm{R}k} & k=\textrm{R}\\
C_{\textrm{LR}} & k=\textrm{F}\\
\sum_{l\in\mathcal{N}_k}C_{kl} & \textrm{otherwise.}
\end{array}
\right.
\]

We model SGs and IBRs as RLC circuits with voltage and/or current sources, and loads as RL circuits. We denote the set of buses with voltage sources as $\mathcal{V}$. We denote voltage sources $v_k^{\circ}(t)$, current sources $i_k^{\circ}(t)$, and load current $\iota_k(t)$.

In the state-space model (\ref{eq:ss}), the state, $x(t)$, consists of the bus voltages, line currents, and load currents. The input, $u(t)$, consists of the voltage and current sources. The matrices $A_m$, $A_m^{\eta}$, and $B$ are specified by the equations in Section~\ref{app:dyn}. The equations in Section~\ref{sec:app:relobs} specify the output matrices, $H^{\eta}$. These matrices also fully specify the incremental model in Section~\ref{sec:incremental}. The structure of the incremental model is detailed in Section~\ref{app:incremental}.


\subsection{Dynamics}\label{app:dyn}

The bus voltage dynamics are
\begin{subequations}
\label{eq:bus}
\begin{align}
i_k^{\circ}(t) + \sum_{l\in\mathcal{N}_k}i_{lk}(t) &= C_k(m)\frac{dv_k(t)}{dt}+\iota_k(t),\quad k\notin\{\mathcal{V},\textrm{F}\}\label{eq:busvolt1}\\
v_k(t)&=v_k^{\circ}(t),\quad k\in\mathcal{V}.\label{eq:busvolt2}
\end{align}
If there is no current source at bus $k$, $i_k^{\circ}(t)=0$. Similarly, if there is no load at bus $k$, $\iota_k(t)=0$. Otherwise, the load current dynamics are
\begin{align}
v_k(t) &=L_k\frac{d\iota_k(t)}{dt}+R\iota_k(t).
\end{align}
At bus $\textrm{F}$, the voltage dynamics before a fault are
\begin{align}
i_{\textrm{LF}}(t) + i_{\textrm{RF}}(t) &= \frac{C_{\textrm{LR}}}{2}\frac{dv_\textrm{F}(t)}{dt}.
\end{align}
Let $G_{\textrm{F}}(m)$ be the fault's conductance matrix. During a fault, the voltage dynamics are
\begin{align}
i_{\textrm{LF}}(t) + i_{\textrm{RF}}(t) &= \frac{C_{\textrm{LR}}}{2}\frac{dv_{\textrm{F}}(t)}{dt} + G_{\textrm{F}}(m)v_{\textrm{F}}(t).
\end{align}
Note that all faults must have some resistance in this formulation, i.e., bolted faults are modeled with a very small resistance.

The current through line $kl\in\mathcal{L}\setminus\{\textrm{LF},\textrm{RF}\}$ evolves as
\begin{align}
v_{k}(t)-v_{l}(t) &= R_{kl}i_{kl}(t)+L_{kl}\frac{di_{kl}(t)}{dt}.\label{eq:unfaultedi}
\end{align}

The currents through either side of the faulted line are
\begin{align}
v_{\textrm{L}}(t)-v_{\textrm{F}}(t) &= 
m_{\textrm{T}}\left(R_{\textrm{LR}}i_{\textrm{LF}}(t)+L_{\textrm{LR}}\frac{di_{\textrm{LF}}(t)}{dt}\right)\label{eq:faultlineLF} \\
v_{\textrm{R}}(t)-v_{\textrm{F}}(t) &= 
(1-m_{\textrm{T}})\left(R_{\textrm{LR}}i_{\textrm{RF}}(t)+L_{\textrm{LR}}\frac{di_{\textrm{RF}}(t)}{dt}\right).
\end{align}
\end{subequations}

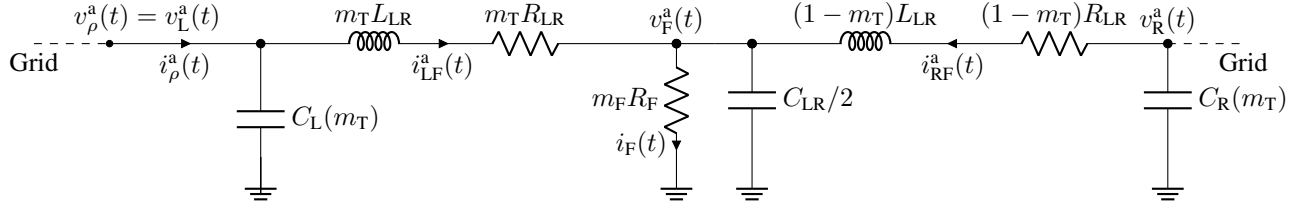
\begin{figure*}[h!]
\begin{center}
\begin{circuitikz}[american currents]
    \draw[dashed] (-3,0)node[below]{Grid} to (-2,0);
    \draw (-2,0)node[above]{\hspace{10mm}$v_{\rho}^{\textrm{a}}(t)=v_{\textrm{L}}^{\textrm{a}}(t)$} to [/tikz/circuitikz/bipoles/length=30pt,short,i_>=$i_{\rho}^{\textrm{a}}(t)$,*-*](0,0);
    \draw[dashed] (12,0) to (13,0)node[below]{Grid};
	\draw (0,0)  to [open, *-*] (5.5,0) 
    node[above]{$v_\textrm{F}^{\textrm{a}}(t)$}
    to [open, *-*] (6.5,0) to [open, *-*] (12,0)
    node[above]{$v_\textrm{R}^{\textrm{a}}(t)$};
    \draw (0,0) to [/tikz/circuitikz/bipoles/length=30pt,C=$C_{\textrm{L}}(m_{\textrm{T}})$] (0,-2)
    to (0,-1.5) node[ground]{};
    \draw (12,0) to [/tikz/circuitikz/bipoles/length=30pt,C=$C_{\textrm{R}}(m_{\textrm{T}})$] (12,-1.5)
    to (12,-1.5) node[ground]{};
	\draw (0,0) to [/tikz/circuitikz/bipoles/length=30pt,L=$m_{\textrm{T}}L_{\textrm{LR}}$,i_>=$i_{\textrm{LF}}^{\textrm{a}}(t)$] (3,0)
    to [/tikz/circuitikz/bipoles/length=30pt,R=$m_{\textrm{T}}R_{\textrm{LR}}$ ] (4,0) to (6,0);
  	\draw (5.5,0) to [/tikz/circuitikz/bipoles/length=30pt,R,l_=$m_{\textrm{F}}R_{\textrm{F}}$,i_>=$i_{\textrm{F}}(t)$] (5.5,-1.5)
    to (5.5,-1.5) node[ground]{};
    \draw (6.5,0) to [/tikz/circuitikz/bipoles/length=30pt,C=$C_{\textrm{LR}}/2$] (6.5,-1.5)
    to (6.5,-1.5) node[ground]{};
    \draw (6,0) to [/tikz/circuitikz/bipoles/length=30pt,L=$(1-m_{\textrm{T}})L_{\textrm{LR}}$,i_<=$i_{\textrm{RF}}^{\textrm{a}}(t)$] (10,0)
    to [/tikz/circuitikz/bipoles/length=30pt,R=$(1-m_{\textrm{T}})R_{\textrm{LR}}$ ] (11,0) to (12,0);
\end{circuitikz}
\end{center}
\caption{Circuit model of line LR during a phase a-to-ground fault. The relay is located at bus L.}
\label{fig:network}
\end{figure*}

\subsection{Relay observation and pre-fault voltage}\label{sec:app:relobs}

Assume the relay is not at an SG bus. Then the relay observes the voltage,
\begin{subequations}
\label{eq:relobs}
\begin{align}
v_{\rho}(t)&=v_{\textrm{L}}(t),
\end{align}
and the current departing bus L to F. This is not $i_{\textrm{LF}}(t)$, which already has the line charging current subtracted off. The current observed by the relay is
\begin{align}
i_{\rho}(t)&=i_{\textrm{LF}}(t) + \frac{1}{2}m_{\textrm{T}}C_{\textrm{LR}}\frac{dv_{\textrm{L}}(t)}{dt}.\label{eq:relaycurrent}
\end{align}
\end{subequations}
We can use (\ref{eq:busvolt1}) to substitute state variables in for the derivative.

Given $v_{\rho}(t)$ and $i_{\rho}(t)$, the relay can compute the current, $i_{\textrm{LF}}(t)$ from (\ref{eq:relaycurrent}). The fault voltage is then given by
\begin{align}
v_{\textrm{F}}(t) &= v_{\rho}(t) - m_{\textrm{T}}\left(
L_{\textrm{LR}}\frac{i_{\textrm{LF}}(t)}{dt}
+R_{\textrm{LR}}i_{\textrm{LF}}(t)
\right)
\end{align}
Computing $v_{\textrm{F}}(t)$ involves two derivatives. This is reasonable given the sampling rates of digital relays; and, in some cases the capacitance might be negligible, eliminating one of the derivatives.

\subsection{Incremental model}\label{app:incremental}

The incremental model is identical to in Appendix~\ref{app:dynamics}, except for the voltage dynamics at bus F:
\begin{align}
\tilde{i}_{\textrm{LF}}(t) + \tilde{i}_{\textrm{RF}}(t) &= \frac{1}{2}C_{\textrm{LR}}\frac{d\tilde{v}_{\textrm{F}}(t)}{dt}+G_{\textrm{F}}(m)\left(\tilde{v}_{\textrm{F}}(t)+v_{\textrm{F}}(t-\delta)\right).
\end{align}

We can compute $v_{\textrm{F}}(t-\delta)$ from $v_{\rho}(t-\delta)$ and $i_{\rho}(t-\delta)$, as described in Appendix~\ref{sec:app:relobs}.

\subsection{Model reduction via singular perturbation}\label{app:singular}

Let $G^{\eta}$ be an invertible matrix for which
\[
G^{\eta}\tilde{x}(t) = \begin{bmatrix}
y^{\eta}(t)\\w(t)
\end{bmatrix},
\]
and
\[
\Gamma^{\eta}_m = G^{\eta}A^{\eta}_m\left(G^{\eta}\right)^{-1},\quad \hat{\Gamma}^{\eta}_m=G^{\eta}\hat{A}_m^{\eta},\quad \beta^{\eta} = G^{\eta}B.
\]
$G^{\eta}$ is a coordinate transformation that makes the relay's observation a part of the state vector, with $w(t)$ representing the rest of the states. Let the subscript $y$ indicate indices corresponding to states $y^{\eta}(t)$ and the subscript $w$ to states $w(t)$. Then
\begin{align*}
\Omega_m^{\eta} &= \Gamma_m^{\eta}[y,y]-\Gamma_m^{\eta}[y,w]\Gamma_m^{\eta}[w,w]^{-1}\Gamma_m^{\eta}[w,y]\\
\hat{\Omega}_m^{\eta} &= \hat{\Gamma}_m^{\eta}[y]-\Gamma_m^{\eta}[y,w]\Gamma_m^{\eta}[w,w]^{-1}\hat{\Gamma}_m^{\eta}[w]\\
\chi_m^{\eta} &= \beta_m^{\eta}[y]-\Gamma_m^{\eta}[y,w]\Gamma_m^{\eta}[w,w]^{-1}\beta_m^{\eta}[w].
\end{align*}

\balance
\bibliographystyle{IEEEtran}
\bibliography{MainBib,JATBib}

\end{document}